\documentclass[a4paper,12pt]{article}
\usepackage{amsmath}
\usepackage{mathtools}
\usepackage{graphicx}
\usepackage{hanging}
\usepackage{authblk}
\usepackage{float}
\usepackage[font={footnotesize}]{caption}

\begin{document}
\title{Predicting Recall Probability to Adaptively Prioritize Study}
\author[1]{Shane Mooney}
\author[2]{Karen Sun}
\author[3]{Eric Bomgardner}
\affil[1]{Quizlet, Inc. \authorcr Email: {\tt shane@quizlet.com}\vspace{1.5ex}}
\affil[2]{Quizlet, Inc. \authorcr Email: {\tt karen@quizlet.com}\vspace{1.5ex}}
\affil[3]{Quizlet, Inc. \authorcr Email: {\tt eric.bomgardner@quizlet.com}\vspace{1.5ex}}
\date{\today}
\maketitle

\section{Abstract}
Students have a limited time to study and are typically ineffective at allocating study time. Machine-directed study strategies that identify which items need reinforcement and dictate the spacing of repetition have been shown to help students optimize mastery (Mozer \& Lindsey 2017). The large volume of research on this matter is typically conducted in constructed experimental settings with fixed instruction, content, and scheduling; in contrast, we aim to develop methods that can address any demographic, subject matter, or study schedule. We show two methods that model item-specific recall probability for use in a discrepancy-reduction instruction strategy. The first model predicts item recall probability using a multiple logistic regression (MLR) model based on previous answer correctness and temporal spacing of study. Prompted by literature suggesting that forgetting is better modeled by the power law than an exponential decay (Wickelgren 1974), we compare the MLR approach with a Recurrent Power Law (RPL) model which adaptively fits a forgetting curve. We then discuss the performance of these models against study datasets comprised of millions of answers and show that the RPL approach is more accurate and flexible than the MLR model. Finally, we give an overview of promising future approaches to knowledge modeling.

\section{Introduction}
Students need to memorize and recall knowledge components (KC) to pass exams in academic and professional settings: this is a requirement that spans nearly all demographics and disciplines. To prepare for these tests, students need to make decisions that maximize the likelihood of recall for all KCs on the test date. Given large swaths of content to learn, students must wade through material and distill source material into lists of digestible KCs. With these components in hand, students need to organize, review, and reinforce them for efficient practice. They must accurately monitor what they know and need to learn, apply the right techniques, and allocate study time appropriately; achieving this has a large impact on test performance  (Ariel, 2013; Kornell \& Metcalfe, 2006; Metcalfe \& Kornell, 2005; Thiede, Anderson, \& Therriault, 2003; Thiede \& Dunlosky, 1999; Pyc \& Dunlosky, 2010).

There are several common methods utilized by students to practice KCs, both self-directed and instructor-directed. Given primary sources or textbooks, students will read and then re-read material as review, sometimes using note-taking to distill facts and key concepts from long-form text. A more structured form of such notes can be realized as flashcards, where students or teachers break concepts into paired text, sometimes including other media like images, to represent knowledge components. Finally, the practice test (or study guide) is a commonly used format for self-testing wherein students practice recalling knowledge in a setting that closely mimics the final exam. These practice tests are typically acquired through a bank made available by a teacher, or in some cases can be purchased through publishers targeting practice for certain standardized exams. The efficacies of these existing forms of study have been studied extensively, showing that testing, as opposed to re-reading or non-generative review, typically has the highest positive effect on future recall probability (Dunlosky, 2013). As a result, there has been much research into developing teaching software that leverages repeated testing as a study tool (Ridgeway, Mozer, \& Bowles 2016; Lindsey, et. al 2014; Ritter et. al 2007; Anderson, Conrad, \& Corbett 1989).

\subsection{Quizlet: an online learning platform}
Quizlet.com is an online learning platform that allows students to encode multimedia knowledge components into digital flashcards. Students and teachers can create lists of flashcards up to any length and use them for self study or share them with other users. Quizlet offers a range of practice experiences for these lists of flashcards, ranging from traditional flashcard review to spatial study games to dynamic practice test generation. One in particular, called \textit{Learn}, generates an adaptive study path on a question-by-question basis, choosing the next KC and question format that should be surfaced at each step to maximize the likelihood of a student's overall mastery of a given list of flashcards. 

As of the end of 2017, about 30 million teachers and students use Quizlet each month to create and/or study flashcards. Over 8 billion flashcards, over 200 million lists, and over 30 billion answers have been logged on the platform, which provides a remarkable dataset for research. The most common learning goal on the platform is foreign vocabulary memorization, followed by preparation for biology and psychology oriented exams; the primary users are U.S. high schoolers enrolled in the public school system. In this paper we discuss the development of a subject matter, language, and demographic agnostic algorithm for allocating a student's study time to optimize their mastery of a given list of flashcards through the \textit{Learn} study experience.

\subsection{Knowledge state modeling}
There are many aspects of the study process that can and should be programmatically optimized. Since students only have limited time to study, the first task is to accurately determine which knowledge components they are least likely to recall during the test. With a notion of strong and weak KCs in hand, the ideal reinforcement technique (whether generative or non-generative, question components, visual or textual mnemonic aids, to give some examples) can be deduced and generated. The ideal study spacing can be derived and individual KCs scheduled for reinforcement. This final aspect of study has been explored extensively, with most literature concluding that spaced study is superior to massed study, and that a student can benefit from scheduling dictated by spaced repetition software or SRS (Cepeda et. al 2009; Cepeda et. al 2008; Cepeda et. al 2006; Cepeda et. al 2006).

Because our population consists largely of high school students studying KCs associated with a textbook or guided classroom instruction, the content selection component of the process is less important than helping students prioritize the order of KCs during study. Additionally, the majority of students use the platform to study the day before a test, so we don't have as much opportunity to leverage a scheduling algorithm. Therefore, this study focuses on the problem of optimizing our platform's ability to predict the recall probability of any KC at any point in time. This would effectively enable us to rank KCs within a list from weakest to strongest, and act on the assumption that the weakest KC should always be the first to be reinforced. This is a greedy optimization since it is locally optimal to attempt to reset the knowledge state of the weakest KC to 1 through feedback, thereby increasing the overall mastery of the list to as much as possible. This strategy is consistent with the discrepancy-reduction mechanism (Dunlosky \& Thiede, 1998), a frequently observed student-directed item allocation strategy during self-paced study. That being said, this is a baseline implementation which does not consider more long-lived outcomes related to list strength effect, failure to reset the knowledge state to 1, or long term memory of a list past the test date. There is ongoing work to model and optimize the other aforementioned components to provide an integrated end-to-end programmatic study advisor.

The task of accurately predicting recall requires a model of a student's knowledge state for any given KC at a certain point in time. Existing mathematical and computational models of knowledge state frequently rely on a model of forgetting, or how the probability of recall for a KC decays over time without reinforcement. There has been extensive research for an accurate forgetting model since Ebbinghaus' self-experimentation and initial establishment of the general nature of memory decay (Ebbinghaus, 1885/1964). Generally, it has been shown that a power law decay explains the data the best (Rubin \& Wenzel, 1996; Wixted \& Carpenter, 2007; Wixted 2004; Wixted \& Ebbesen, 1991; Wixted \& Ebbesen 1997), while exponential decay as the kernel function for forgetting has evolved into a nearly exponential function through the contextual drift framework (Mensink \& Raaijmakers, 1988; Murdock, 1997; Howard \& Kahana, 2002). The declarative knowledge module of the ACT-R cognitive architecture model (Anderson et al., 2004) essentially assumes a power law decay for traces laid down by reinforcements of each KC. These theories are subject matter agnostic and are elegant, interpretable baselines for understanding item-specific knowledge state. 

	Another framework that provides a model for knowledge state as well as likelihood of correct question application is Bayesian Knowledge Tracing or BKT (Corbett, Anderson 1995), which postulates that the internal knowledge state is incrementally affected by the previous answer given; subsequently, the ability to answer a question is influenced by both a guessing and slippage parameter. Free parameters associated with each of the models mentioned can be derived from a suitable amount of training data and gives models the flexibility of personalization such as rate of memory decay in the forgetting curve and slippage for BKT.

Most of these studies have drawn conclusions and compared knowledge state models amongst populations of subjects that have their demographics, instruction, content, and scheduling regimented by the experiment. Both for practical and scientific purposes, we wanted to demonstrate a knowledge model that could cut across demographics, subject matter, and study schedule. 

\section{Methodology}
Our starting point was evidence from research that knowledge state is at the very least a function of practice history and time elapsed since a previous calculation. We began with an application of a traditional machine learning approach, multiple logistic regression (MLR), to produce a portable and inspectable model and used learnings to subsequently create a custom forgetting algorithm, the Recurrent Power Law model (RPL). 

\subsection{Multiple Logistic Regression}
Multiple Logistic Regression (MLR) is a simple method for predicting categorical variables, in our case whether or not a student will be able to recall an item at a given point in time. An item corresponds to a "side" of a flashcard containing text, with the potential for either item of a KC to act as a cue for the other during cued recall. MLR produces a very interpretable model, generating weights for independent explanatory variables. Prior research points to several strong predictors, including success or failure to recall the KC on previous trials (Corbett \& Anderson, 1995), time elapsed since last trial (Cepeda et. al 2008), time between previous trials (Cepeda et. al 2008; Wixted \& Carpenter, 2007), the format of previous trials , and the \textit{direction} of past study (i.e. recalling A given B vs. recalling B given A) (Kahana \& Caplan, 2002; Murdock 1966). MLR also produces a fast, lightweight model, which makes it ideal for implementation in an online teaching system at Quizlet's scale that must make predictions for every student, for every KC, and update those predictions adaptively as items are studied.

All predictors for our model are derived from past trials by a specific student on a specific KC. The strongest predictors are derived from the sequence of successes or failures on the student's past trials on the KC. Because more recent trials are more predictive of future recall, we assign indices to past trials in reverse chronological order. Trial 1 (if any) being the most recent, up to some limit $n$. This threshold was determined experimentally by increasing increasing this value until it no longer produced statistically significant weights at $p=0.05$-- refer to the Appendix for more details.

\begin{equation}
a = \sum_{i=1}^n w_{c_i} c_i
\end{equation}

$c_i$ is $0$ or $1$ depending on whether the trial was incorrect or correct, respectively.

Forgetting is assumed to follow an approximately exponential decay of an independent trace corresponding to each previous trial. Again ranking previous trials in reverse chronological order, we find the number of seconds elapsed between the current time and each previous trial, and take the natural log of this value to linearize the underlying exponential decay. A weight is assigned to each log-time delta where $t$ is the time of the trial at that index.

\begin{equation}
b = \sum_{j=1}^m w_{t_j} ln(t_{current}- t_j)
\end{equation}

The window size $m$ determining the number of trials to look back is again determined experimentally.

We assume a spacing effect proportional to the natural log of the time between the two most recent trials, if any.

\begin{equation}
c = \sum_{k=1}^{l} w_{s_k} ln(t_{k+1} - t_{k+2})
\end{equation}

Again, we experimentally determined the window size $l$.
	
While we've thus far treated study in either direction as equivalent, past study in the current direction should be more strongly predictive of successful recall than study in the opposite direction. We model this by accounting for the proportion of past trials on the KC that were in the current direction.

\begin{equation}
d = w_{r_0} \frac{count(trial_s)}{count(trial)}
\end{equation}

where $trial_s$ are the trials that occurred on the same side of the flashcard and a $trial$ occurrence is any trial.
	
Finally, we introduce two features meant to capture information from trials beyond the limited windows of recent trials we arrived at above:

\begin{equation}
e = w_{r_1} count(trial) + w_{r_2} ln(t_{current} - t_{firstTrial})
\end{equation}

The final assembled equation to predict recall probability is:

\begin{equation}
\begin{split}
logit[\pi(t_{current})] = \beta + a + b + c + d + e
\end{split}
\end{equation}
	
where $\pi(t)$ is the probability of recall at current time $t$.

We trained this model on approximately 1.5 million questions from Quizlet's \textit{Write} instruction system, which given a user-generated set of KCs and user-selected study direction, prompts the student with cued recall questions. To answer a question correctly, a student must type the answer to the cue accurately character for character except for some punctuation-based edge cases. The \textit{Write} system will repeat incorrectly recalled KCs until all KCs are recalled correctly, at which point the student may choose to repeat the process.
	
The model was trained using the Newton-Raphson method to find a maximum likelihood estimate of the parameters. The resulting coefficients indicated that the history of correctness across trials were the most predictive variables, followed by the proportion of same side trials. More recent trial time spacings were more predictive than older ones, as expected. Refer to the Appendix for a comprehensive overview of our empirically derived coefficients. 

\subsection{Recurrent Power Law}
The MLR model's use of a log of time between trials implies an exponential decay forgetting curve assumption. Literature indicates that power law decay better describes forgetting than exponential decay (Wixted \& Carpenter 2007). Additionally, MLR is limited in its assumption that each feature has an independent, linear effect on the log-odds of the recall probability. This limitation made it difficult to account for differences in question formats, which have a non-obvious and non-linear relationship with other model features.

To address these limitations, we developed a model that explicitly fits a series of power law forgetting curves to each series of trials on a KCs. Recall probability is reset to $1$ after each trial because incorrect answers are immediately followed with feedback informing the student of the correct answer. Recall probability then follows a power law decay with a shape that is successively updated based on the outcome of each trial. Because the model at its core relies on a power law decay function, and because the parameters of that function are recurrently updated with each trial, we refer to it here as a Recurrent Power Law (RPL) model.

\begin{figure}[H]
\centering
\includegraphics[width=0.8\textwidth]{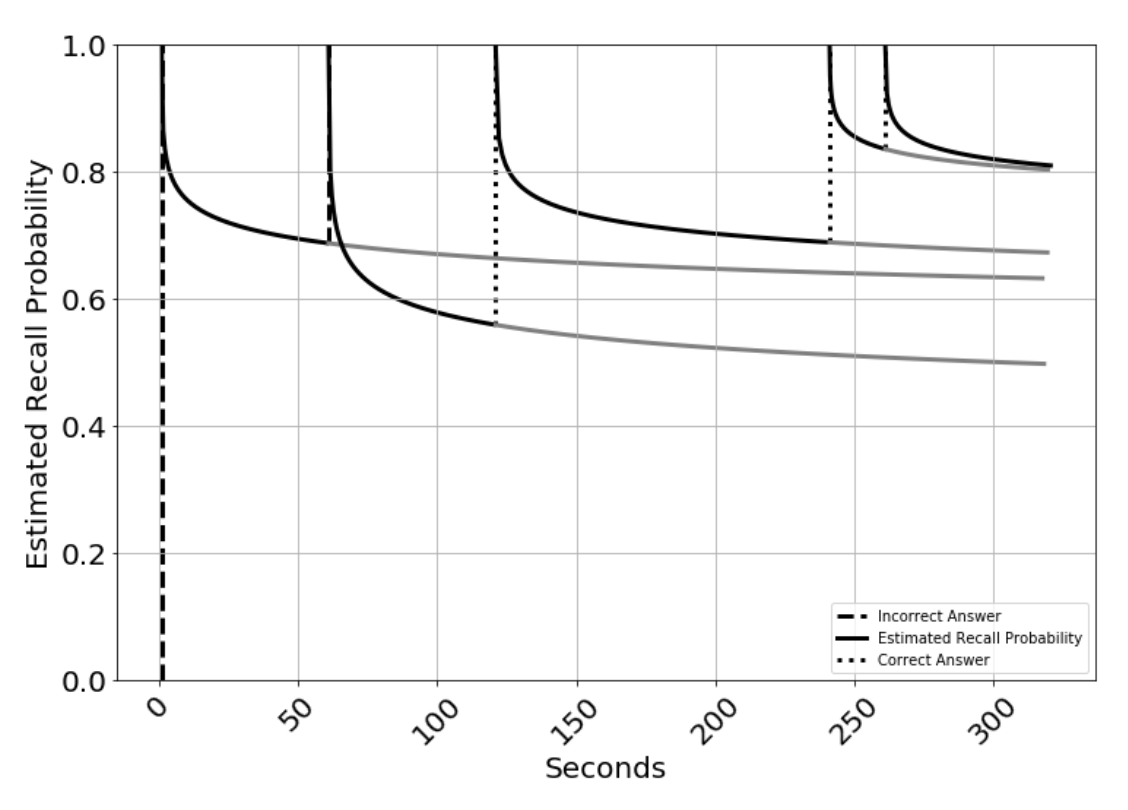}
\caption{Example of estimated forgetting curves following power law decays as generated by the RPL model. Forgetting curves get steeper after incorrect answers and shallower after correct answers. Estimated recall probability always resets to 1 after each question because Quizlet always gives the user immediate feedback. }
\end{figure}

The general shape of the forgetting curve is described by:

\begin{eqnarray}
s\prime = e^{-s} \\
\tau\prime = e^{-\tau}  \\
p_{cr} = (1 + s\prime r)^{-\tau\prime}
\end{eqnarray}

	where $r$ is the retention interval in seconds, and $s$ and $\tau$ are fit parameters describing the shape of the forgetting curve. $p_{cr}$ is interpretable as recall probability for a cued response question.
	
	On the first trial, $s$ is set to its initial value, $s_0$. $\tau$ is initialized based on the outcome of the first trial, $\tau_{0c}$ if correct, and $\tau_{0i}$ if incorrect. On subsequent trials, $s$ and $\tau$ are updated based on a function of their previous values, the current estimated cued response recall probability $p_{cr}$, the assumed guess probability $g$, the outcome of the trial, and additional fit parameters $\gamma_c$, $\gamma_i$, $\tau_c$, $\tau_i$, $s_c$, and $s_i$ modulating the magnitude of the update:
	
\begin{equation}
\begin{split} 
\gamma, \tau, s 
\begin{cases}
    \begin{split}
    \hspace{20 mm} \\
    \gamma = \gamma_c \\
    \tau = \tau_{n - 1} ( 1 + ( \tau_c (1 - p_{cr}) ) ^ \gamma ( 1 - g_f ) ) \\
    s = s_{n - 1} ( 1 + s_c (1 - p_{cr}) ( 1 - g_f ) ) \\
    \hspace{20 mm}
    \end{split}
    ,& \text{if } c_n = 1\\
    \begin{split}
    \hspace{20 mm} \\
    \gamma = \gamma_i \\
    \tau = \tau_{n - 1} ( 1 - \frac{( \tau_i p_{cr} ) ^ \gamma}{ 1 - g_f }) \\
    s = s_{n - 1} ( 1 - \frac{s_i p_{cr}}{1 - g_f} )\\
    \hspace{20 mm}
    \end{split}
    ,& \text{if } c_n = 0
\end{cases}
\end{split}
\end{equation}
	
Note that in the case of successful recall, when $p_d$ is near $1$, corresponding to short retention intervals, the updates to $\tau$ and $s$ are small, while at longer retention intervals when $p_d$ is small, the updates to $\tau$ and $s$ are larger. This models the spacing effect as a function of recall probability, consistent with the retrieval effort hypothesis (Pyc \& Rawson 2009). Inversely, in the case of failed recall, the updates to $\tau$ and $s$ are larger at shorter retention intervals, when $p_d$ is large. The parameter $\gamma$ allows this relationship to have a non-linear form.
	
The parameter $g_f$ serves to reduce the magnitude of change to the forgetting curve parameters $\tau$ and $s$ in instances where a question may have been answered correctly purely by chance in the case that the student can't actually recall the correct answer. We fix $g_f$ to $1$ divided by the number of options available for the given question format, for example 0.25 for a 4-option multiple choice question and 0.5 for a true/false question. For cued recall, we assume no chance of guessing by chance $(g_{cr}=0)$. When $g_f$ is larger, the updates to $\tau$ and $s$ in the case of a correct answer are small, because correct answers are not guaranteed to indicate accurate recall.

Even after accounting for the possibility of guessing by chance, the underlying probability of a student knowing the answer to a question varies by question type. This can likely be attributed to the requirement to either recall an item (cued recall), or to merely recognize it (multiple choice). For each non-cued recall question format, we fit a "difficulty factor" $k$, corresponding to a ratio of the odds that a student knows the answer to a question of the given format (for the given item), divided by the odds that a student would know the answer when prompted as a cued recall question.

\begin{equation}
p_k= \frac{k_f p_{cr}}{1 - p_{cr}(1 - k_f) }
\end{equation}

The probability of a correct answer is a function of the probability of guessing $g_f$ and the probability of knowing the answer, given the question format $f$:

\begin{equation}
p_c=p_k + (1 - p_k) g_f
\end{equation}

Note that as $p_k$ approaches $1$, the influence of $g_f$ approaches $0$. This implies that the probability of guessing becomes small as the probability of knowing the answer gets larger. This formulation of guessing is reminiscent of Bayesian Knowledge Tracing (Corbett \& Anderson, 1995).

Finally, we must account for study direction. We treat the two possible studyable directions of a two dimensional KC (recalling A when cued with B vs recalling B when cued with A) as two separate but interacting items. Each item maintains its own $\tau$ and $s$ parameters, updated as described above, and only when the KC is studied in that direction. However, if there exists at least one past trial of the inverse item, we assume the possibility that the target item was forgotten (or never learned), but that inverse item is both remembered and that knowledge transfers to the target item.

\begin{equation}
p=p_c + (1 - p_c)(p_o t)
\end{equation}
Where $p_o$ represents the $p_c$ value for the inverse item, and $t$ represents a fixed probability of inverse item transfer.

The model was trained using the Nelder-Mead method to find a maximum likelihood estimate of the parameters. Refer to the Appendix for more detail about the parameters we arrived at.

\section{Results}
Both models were trained on data collected from usage of two of Quizlet's instruction systems. The first is Quizlet's\textit{Write}, an instruction system that uses only cued recall questions. The second is Quizlet's \textit{Learn}, an instruction system that incorporates cued recall, multiple choice, and self-graded (know or don't know) flashcard questions. Quizlet \textit{Learn} also incorporates a knowledge model into its algorithm for prioritizing KCs for study. Subject to other constraints, the algorithm attempts to greedily schedule the KCs predicted to be least likely to be recalled correctly. Here we compare results for a version of \textit{Learn} that uses the RPL model for this purpose.

\begin{figure}[H]
\centering
\includegraphics[width=0.8\textwidth]{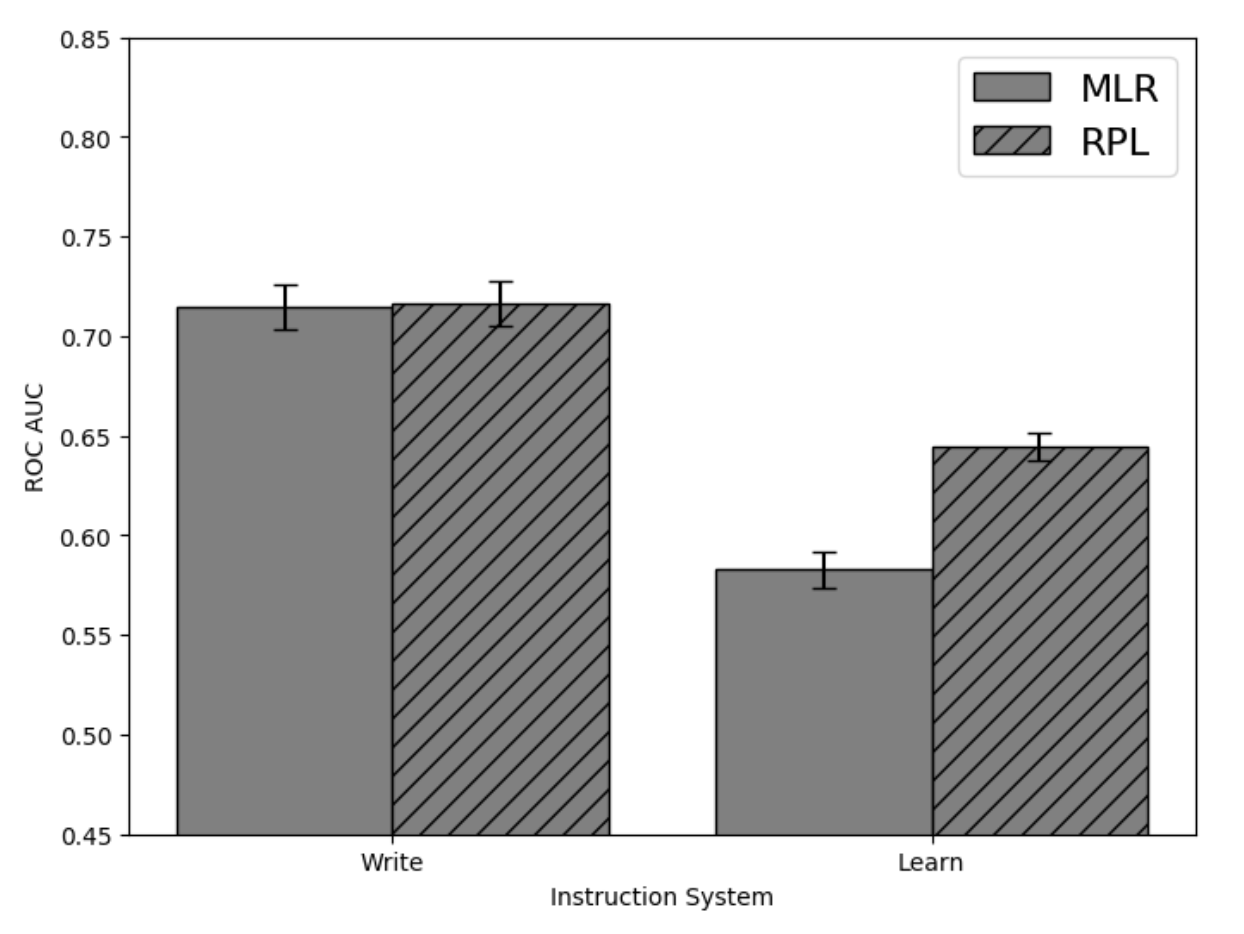}
\caption{AUC scores for each model for Write (cued recall only) and Learn (mixed question format). Error bars indicate $\pm$ 1.96 standard error of the mean. The introduction of question formats other than cued recall makes the knowledge modeling problem more difficult, but RPL better handles this case.}
\end{figure}
	
Both models do well for \textit{Write}, with only cued recall questions, and performance of both suffers for \textit{Learn}, which introduces variable question formats. However, for \textit{Learn} with a variety of question formats, RPL significantly outperforms MLR in this case. While MLR is unable to capture the impact of differences in question format, RPL is able to do so, making it a more general model that's not limited to modeling a single type of study.

Since past study history on a KC is the only input, both models perform much better in cases where there exist at least two past study trials on the KC.

\begin{figure}[H]
\centering
\includegraphics[width=0.8\textwidth]{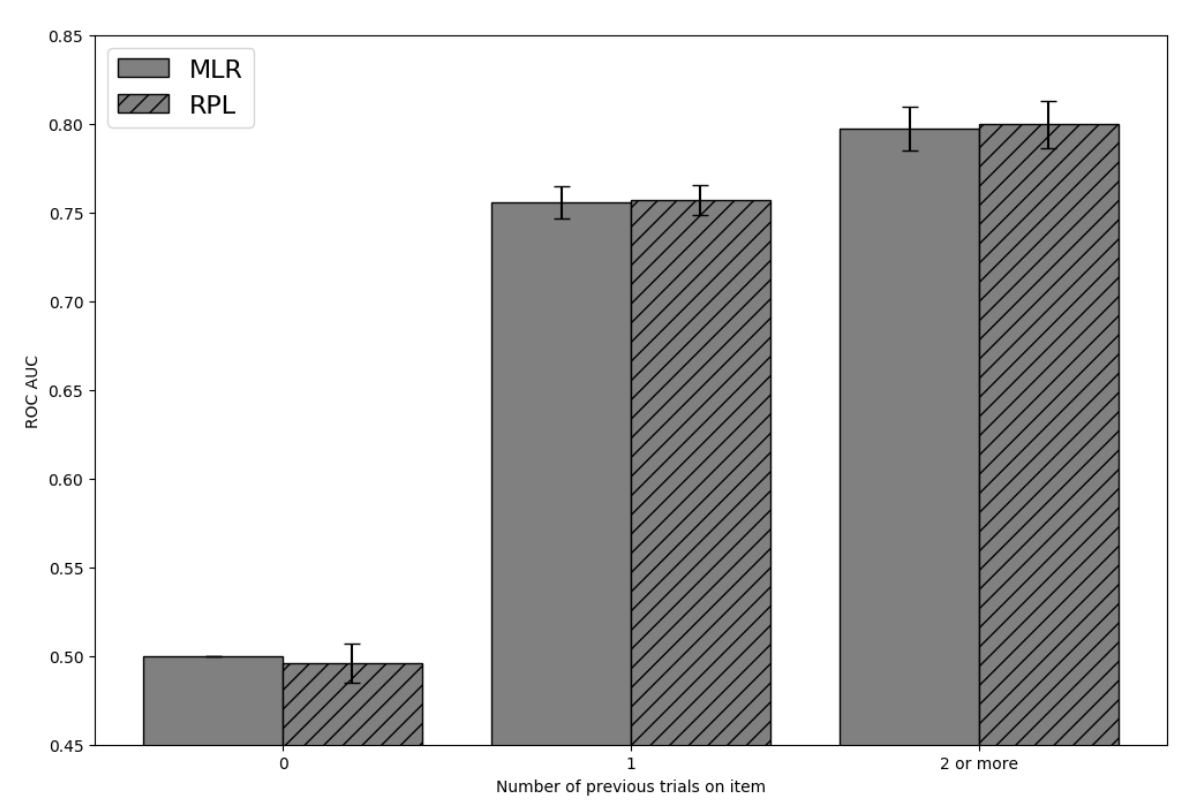}
\caption{AUC scores for each model for Write (cued recall only), segmented by number of past trials on the KC. Error bars indicate $\pm$ 1.96 standard error of the mean. Performance improves significantly when there are at least two past trials on each KC. Trying to predict the first answer is on par with random guessing which is to be expected since there is no context on the student's knowledge state.}
\end{figure}

\section{Discussion}
The RPL model integrates mechanisms from many existing knowledge state models including the power law decay of memory, contextual traces from cued recall of both items of a KC, guessing, and an odds ratio that encapsulates aspects of question format difficulty. It has been a practical success-- it is inspectable, simple to reason about, portable, and content agnostic but question format aware. However, there are still many parameters that could be encoded, including but not limited to intrinsic item difficulty, the topic being learned (e.g. vocabulary, cell division lifecycle, historical battles), and adaptive modeling of a student's pre-existing knowledge state based on initial question performance. Intrinsic item difficulty has been especially explored in foreign language acquisition, with findings that characteristics of word pairs like cognates or grouping target words by valence and concreteness have a significant impact on memory retention (Kroll \& Tocowicz, 2005). Furthermore, we have a large volume of behavioral data from which we can deduce which items have been historically harder to acquire than others, implying that we could perhaps arrive at item-specific difficulty weights empirically. A necessary practical concern for any digital tutor is the reality that students receive instruction in other environments, so establishing the new baseline of knowledge each time a student begins a new study session is essential for efficient item scheduling.

The extensions mentioned above address content-aware modeling, but not of the individual student. Since our models were globally trained, the decay and update parameters for every student was the same, an aggregate across all individuals. Instead, given enough trial data for any individual, we could adapt parameters such as the rate of forgetting to be personalized. Personalized teaching software has also seen success in the field such as with the system developed by Lindsey et. al (2014). 

Beyond the scope of RPL improvements, and towards the goal of an end-to-end adaptive study experience, we need to move beyond two dimensional cue-target KC encoding and implement more comprehensive KC modeling and question generation. For any given knowledge component, there commonly exists on the order of 5-10 popular cues: a foreign vocabulary word could be cued by the translation in the native language, the spoken audio, a representative picture, the definition in the native or foreign language, an example sentence in the target language with the word omitted, or a synonym. These are valuable devices that could be used to generate more optimal questions that better reinforce a student's ability to recall and apply KCs (Murdock, 1997). Our existing models are not robust to these kinds of increases in dimensionality, so there is a need for frameworks that are similarly portable and accurate. On that front, we are exploring promising techniques such as Bayesian Networks in which knowledge states are latent variables, and neural networks which are more suited to modeling a high number of variables with unknown interactions.

\section*{Appendix}

\subsection*{MLR trial lengths}
We experimentally determined the trial lengths (variables $m$, $n$, and $l$) to be the maximum values that still produced statistically significant coefficients on the training data with $p = 0.05$. For our training data, we arrived at $n = 6$, $m = 5$, and $l=3$.

\subsection*{MLR coefficients}

These were coefficients we obtained from training and deployed in production with the MLR model. Notably, the most recent trial correctnesses were the most predictive.

\begin{table}[H]
\centering
\begin{tabular}{|l|l|l|l|l|l|l|}
\hline
\textbf{Coef} & \textbf{Value} & \textbf{Std Err} & \textbf{$z$} & \textbf{$P > |z|$} &\textbf{ 0.025 }& \textbf{0.975 }\\
\hline
$\beta$ & 0.4742 & 0.022  & 21.212  & 0.000  &  0.430  & 0.518 \\
\hline
$w_{c_1}$ & 1.8193 & 0.049 & 37.116 & 0.000 & 1.723 & 1.915\\
$w_{c_2}$ & 0.8491 & 0.065 & 13.077 & 0.000 & 0.722 & 0.976 \\
$w_{c_3}$ & 0.7068 & 0.082 & 8.669 & 0.000 & 0.547 & 0.867 \\
$w_{c_4}$ & 0.4325 & 0.097 & 4.470 & 0.000 & 0.243 & 0.622 \\
$w_{c_5}$ & 0.4940 & 0.112 & 4.409 & 0.000 & 0.274 & 0.714 \\
$w_{c_6}$ & 0.3857 & 0.122 & 3.167 & 0.002 & 0.147 & 0.624 \\
\hline
$w_{t_1}$ &  -0.0958 & 0.011 & -8.869 & 0.000 & -0.117 & -0.075 \\
$w_{t_2}$ &  -0.0778 & 0.011 & -7.153 & 0.000 & -0.099 & -0.057 \\
$w_{t_3}$ & -0.0535 & 0.011 & -4.888 & 0.000 & -0.075 & -0.032 \\
$w_{t_4}$ & -0.0257 & 0.012 & -2.128 & 0.033 & -0.049 & -0.002 \\
$w_{t_5}$ & -0.0238 & 0.008 & -2.849 & 0.004 & -0.040 & -0.007 \\
\hline
$w_{s_1}$ &  0.0657 & 0.012 & 5.264 & 0.000 & 0.041 & 0.090 \\
$w_{s_2}$ & 0.0285 & 0.012 & 2.351 & 0.019 & 0.005 & 0.052 \\
$w_{s_3}$ & 0.0325 & 0.014 & 2.382 & 0.017 & 0.006 & 0.059 \\
\hline
$w_{r_0}$ &  0.2126 & 0.054 & 3.916 & 0.000 & 0.106 & 0.319 \\
$w_{r_1}$ &  -0.0719 & 0.009 & -7.750 & 0.000 & -0.090 & -0.054 \\
$w_{r_2}$ & 0.0407 & 0.011 & 3.618 & 0.000 & 0.019 & 0.063 \\
\hline
\end{tabular}
\caption{Coefficients, errors, and z-statistics for the MLR model}
\end{table}

\subsection*{RPL fit parameters}
These tables are the parameters we arrived at for the RPL model. The $k$ factors are question format dependent and will vary depending on the nature of how difficult it is to answer the question by guessing.

\begin{table}[H]
\centering
\begin{tabular}{|l|l|}
\hline
\textbf{Param} & \textbf{Value} \\
\hline
$s_0$ & -3.51706760045 \\
$s_c$ & 0.00643324313615 \\
$s_i$ & -0.0544722896411 \\
$\tau_{0c}$ & 3.86991863068 \\
$\tau_{0i}$ & 3.54103122648 \\
$\tau_c$ & 0.396606246542 \\
$\tau_i$ & 0.294149151118 \\
$\gamma_c$ & 0.887589628199 \\
$\gamma_i$ & 1.39704082213 \\
$p_0$ & 0.378245635733 \\
\hline
\end{tabular}
\caption{Fit parameter values for RPL model}
\end{table}

\begin{table}[H]
\centering
\begin{tabular}{|l|p{6cm}|l|}
\hline
\textbf{Question Type} & \textbf{Description} & \textbf{Value}\\
\hline
Multiple Choice & Cued recall with four possible answers & 2.055274 \\
\hline
Multiple Choice w/ None & The above question type with a ``none of the above'' option & 1.826852 \\
\hline
True/False & Whether a cue/target item pair is correct associated & 1.9616543 \\
\hline
\end{tabular}
\caption{Difficulty factor $k$ values for example question formats for RPL model}
\end{table}

\section*{Acknowledgements}
Many thanks to Dustin Stansbury, Michael Mozer, Diana Chien for reviewing this manuscript, and all the people who use Quizlet and enable our learning research!

\section*{References}
\begin{hangparas}{.25in}{1}
Anderson, J. R., Conrad, F. G., \& Corbett, A. T. (1989). Skill acquisition and the LISP tutor. Cognitive Science, 13 , 467 - 506.

Benjamin, A. S., \& Tullis, J. (2010). What makes distributed practice effective? Cognitive Psychology, 61 , 228 - 247.

Cepeda, N. J., Coburn, N., Rohrer, D., Wixted, J. T., Mozer, M. C., \& Pashler, H. (2009). Optimizing distributed practice: Theoretical analysis and practical implications. Journal of Experimental Psychology, 27 56 , 236 - 246. 

Cepeda, N. J., Pashler, H., Vul, E., \& Wixted, J. T. (2006). Distributed practice in verbal recall tasks: A review and quantitative synthesis. Psychological Bulletin \& Review, 132 , 364 - 380. 

Cepeda, N. J., Pashler, H., Vul, E., Wixted, J. T., \& Rohrer, D. (2006). Distributed practice in verbal recall tasks: A review and quantitative synthesis. Psychological Bulletin, 132 , 354 - 380. 

Cepeda, N. J., Vul, E., Rohrer, D., Wixted, J. T., \& Pashler, H. (2008). Spacing effects in learning: A temporal ridgeline of optimal retention. Psychological Science, 19 , 1095 - 1102.

Cohen, M.S., Yan V.X., Halamish V., \& Bjork R.A. (2013). Do students think that difficult or valuable materials should be restudied sooner rather than later? J Exp Psychol Learn Mem Cogn. 2013 Nov; 39(6) : 1682 - 96.

Corbett, A. T. \& Anderson, J. R. (1995). Knowledge tracing: Modeling the acquisition of procedural knowledge. User Modeling and User-Adapted Interaction, 4, 253-278.

Dunlosky, J., Rawson, K., Marsh, E., Nathan, M., \& Willingham, D. (2013). Improving students' learning with effective learning techniques: Promising directions from cognitive and educational psychology. Psychological Science in the Public Interest, 14 (1), 4-58.

Dunlosky J. \& Thiede KW. 1998. What makes people study more? An evaluation of factors that affect people's self-paced study and yield ``labor-and-gain'' effects. Acta Psychol. 98:37 - 56

Ebbinghaus, H. (1885/1964). Memory: A contribution to experimental psychology. Mineola, NY: Dover

Howard, M. W. \& Kahana, M. J. (2002). A distributed representation of temporal context. Journal of Mathematical Psychology, 46.

Kahana, M.J. \& Caplan, J.B. Memory \& Cognition (2002) 30: 841.

Kornell N, Bjork RA. The promise and perils of self-regulated study. Psychonomic Bulletin Review. 2007;14:219 - 224.

Kornell N, Bjork RA. Optimising self-regulated study: The benefits - and costs - of dropping flashcards. Memory. 2008;16:125?136. 

Kornell N, Metcalfe J. Study efficacy and the region of proximal learning framework. Journal of Experimental Psychology. Learning, Memory, and Cognition. 2006;32:609 - 622.

Kroll, J. F., and Tokowicz, N. (2005). Models of bilingual representation and
processing: looking back and to the future, in Handbook of Bilingualism:
Psycholinguistic Approaches, eds J. F. Kroll and A. M. B. de Groot (New York,
NY: Oxford University Press), 531 - 533.

Lindsey, R. V., Shroyer, J. D., Pashler, H., \& Mozer, M. C. (2014). Improving students' long-term knowledge retention through personalized review. Psychological Science, 25, 639 - 647.

Mensink, G.-J. M. \& Raaijmakers, J. G. W. (1988). A model for interference and forgetting. Psychological Review, 95, 434 - 455.

Metcalfe J, Kornell N. The dynamics of learning and allocation of study time to a region of proximal learning. Journal of Experimental Psychology: General. 2003;132:530 - 542. 

Metcalfe J, Kornell N. A region of proximal learning model of study time allocation. Journal of Memory and Language. 2005;52:463 - 477. 

Mozer, M. C., \& Lindsey, R. V. (2016). Predicting and improving memory retention: Psychological theory matters in the big data era. In M. Jones (Ed.), Big data in cognitive science. Sussex, UK: Psychology Press.

Mozer, M. C., Pashler, H., Cepeda, N., Lindsey, R. V., \& Vul, E. (2009). Predicting the optimal spacing of study: A multiscale context model of memory. In Y. Bengio, D. Schuurmans, J. Lafferty, C. Williams, \& A. Culotta (Eds.), Advances in Neural Information Processing Systems (Vol. 22, p. 1321-1329).

Murdock, B. B. (1966). Forward and backward associations in paired associates.Journal of Experimental Psychology,71, 732 - 737.

Murdock, B. B. (1997). Context and mediators in a theory of distributed associative memory (TODAM2). Psychological Review, 104, 839 - 862

Pavlik, P., \& Anderson, J. (2008). Using a model to compute the optimal schedule of practice. Journal of Experimental Psychology: Applied, 14, 101 - 117.

Potts R, Shanks DR. (2013) The benefit of generating errors during learning. J Exp Psychol. 2014;143:644 - 67.

Pyc M. A., Dunlosky J. (2010). Toward an understanding of students' allocation of study time: why do they decide to mass or space their practice? Mem. Cogn. 38 431 - 440. 

Ridgeway, K., Mozer, M. C., \& Bowles, A. (2016). Forgetting of foreign-language skills: A corpus-based analysis of online tutoring software. Cognitive Science Journal.

Ritter, S., Anderson, J. R., Koedinger, K. R., \& Corbett, A. (2007) Cognitive Tutor: Applied research in mathematics education. Psychonomic Bulletin \& Review, 14, 249-255.

Rubin, D. C., \& Wenzel, A. E. (1996). One hundred years of forgetting: A quantitative description of retention. Psychological Review, 103, 734 - 760. 

Wickelgren, W. A. (1974). Single-trace fragility theory of memory dynamics. Memory \& Cognition, 2, 775 - 780. 

Wixted, J. T. (2004). On common ground: Jost's (1897) law of forgetting and Ribot's (1881) law of retrograde amnesia. Psychological Review, 111(4), 864 - 879. 

Wixted, J. T., \& Carpenter, S. K. (2007). The Wickelgren power law and the Ebbinghaus savings function. Psychological Science, 18, 133 - 134. 

Wixted, J. T., \& Ebbesen, E. (1991). On the form of forgetting. Psychological Science, 2, 409 - 415. 

Wixted, J. T., \& Ebbesen, E. B. (1997). Genuine power curves in forgetting: A quantitative analysis of individual subject forgetting functions. Memory \& Cognition, 25(5), 731 - 739.
\end{hangparas}
\end{document}